# Empowering Library Users: Creative Strategies for Engagement and Innovation

Snehasish Paul[1*] Shivali Chauhan[2] Atul Kumar Pal[3]

[1]Librarian, Global Institute of Technology & Management, Gurugram, Haryana
[2]Intern, Central Library, Jiwaji University, Gwalior, Madhya Pradesh
[3]Assistant Librarian, Central Library, DULS, University of Delhi

ORCID- https://orcid.org/0009-0003-2730-5314
*Corresponding author email: snehasishpaulas98@gmail.com

## Abstract

This study investigated the integration of cutting-edge technologies and methodologies for creating dynamic, user-centered library environments. In creative strategies for engagement and innovation, library users must be empowered to undertake the new role of modernizing library services and enhancing user experiences. It also enhances the information management and user engagement. This can be attained from personalized approaches, such as recommendation systems to interactive platforms that will have effective experiences tailored to users of different natures. It investigates the consumer engagement practices of enthusiasm, sharing, and learning about their roles in cognitive, affective, and behavioural engagements. Combined, these new approaches will help promote learning, interaction, and growth, add value, and have a more positive impact on users. The challenge for libraries in this rapidly changing, technologically advancing, and digitally networked world, with a base of expectant users, is to remain relevant and engaging. This study discusses innovative strategies for empowering library users and enhancing their engagement through creative and technological approaches. This investigation was conducted to integrate cutting-edge technologies and methodologies into creating dynamic library settings that are user-centered and foster learning, interaction, and personal growth.

# 1. Introduction

Libraries' multifaceted approaches enrich their core services and address shifting community demands. Creative features and innovative programs are central to transforming learning environments into dynamic hubs, for active engagement and collaborative development. Modern libraries have evolved from simple book lending to active learning, innovative, and community-engaging entities. Today, the engagement of creativity and innovative approaches will help organize activities in a way that best serves modern times. The following paper explores different ways to increase user interaction with, and satisfaction from, libraries, where the most important are the integration of advanced technologies and individually tailored services. Libraries can redefine their roles as vibrant centers of innovation and scholarly excellence. This study proposes the establishment of a comprehensive research center within an academic library to transform it into a dynamic hub by integrating advanced resources, expert support, focused training programs, and a robust rewards system. Equipped with state-of-the-art facilities and research managers, not to mention collaborative spaces at one's disposal, it caters to the complex needs of modern research, encourages intradisciplinary collaboration, and accelerates academic discovery. The research will consider the advantages, ways of implementation, and difficulties involved in establishing such a center, and set out the blueprint guiding institutions in the quest to develop stronger research capabilities that fuel breakthrough scholarship. A state-of-the-art research center within an academic library is expected to fill the gaps in specialized research environments through advanced resources, expert support, and collaborative spaces. This paper discusses its benefits, implementation strategies, and challenges, and provides hands-on guidance for any institution seeking to upgrade its research capacities of their institution and engender innovation. In this era of information access, it is challenging for libraries to move beyond their traditional functions to real community hubs of engagement and innovation. From a repository for books and publications, libraries have grown into leading drivers of creativity, research, and lifelong learning. Suppose that they remain relevant in a new vibrant scene. In this case, creative ways will have to be found not only in user engagement but also in facilitating innovation and academic excellence.

## 2. Literature reviews

This literature review aims to change the fast-moving landscape by synthesizing key insights and underexplored avenues within the field through a critical analysis of available research to shed light on current trends and future research opportunities.

**Otike et al. (2022)** discussed innovation strategies in an academic library using entrepreneurial theories; they needed to either adapt themselves or innovate. To this end, both the Competing Values Framework and the Disruptive Innovation Theory have been developed through an argument on the integration of innovative strategies for user engagement and creativity. These theories provide justification for transforming academic libraries into value-adding organizations and therefore support their sustainability and relevance. The authors have supported the fact that business entrepreneurial theories can be used within an academic library to introduce new strategies with efficiency to respond to new or evolving demands and ensure its long-term survival. **Zou et al. (2020)** discussed the social media engagement strategies applied in libraries and pointed out their potential to improve user interaction and participation. This study highlighted three major contributions to the library system, mainly designing strategies for user engagement based on user preferences. By analyzing user data, the authors argue that libraries can design services that are more effective and participatory through social media. This will not only raise the level of user engagement but also align library services with current trends in digital communication and eventually make libraries more interactive and responsive. **Giannaros et al. (2020)** recommended the use of innovative technologies such as recommendation systems, gamification, and indoor localization to enable library services to generate innovative and personalized experiences for visitors. They show how modern libraries can be tailor-made through intelligent ways recommendations, how the gamified features of the place make the place relevant to users regarding its availability, and indoor interactivity through the interactive changes in features within the spaces in the library. This means that with such technologies, further developments will enable libraries to tell who their loyal visitors are, which will help them support such visitors and enhance their interest in the innovation of library services. **Henkel et al. (2018)** discuss open innovation in libraries. Considerable attention has been devoted to users' active involvement in innovation processes, soliciting feedback, and adapting services to their needs. The survey results and six library case studies provide good examples of open innovation and implementation of networked governance strategies. Drawing on these case studies and an online questionnaire, this study discusses how libraries can increase engagement and drive innovation through small- and large-scale initiatives alike, emphasizing the role of information sharing and user involvement in powering library innovation. **Huang et al. (2017)** investigated Twitter engagement strategy for public libraries, which were identified with the help of topic-modeling techniques: literature exhibits, engaging topics, community building, and library showcasing. Thus, this study reflects how social media as a whole and Twitter, in particular, have changed the concept of a library concerning user engagement. This effect is assessed not only by analyzing feedback from users, but also by stressing that change must instead be brought about by users themselves, since only then will it be delivered directly to those concerned, thereby improving the services provided by any library to respond to evolving demands and achieve long-term success.

### 3. Objectives of the study

1. To design and implement a comprehensive research center within an academic library integrating advanced resources and expert support.

2. To develop a structured support system with research managers to assist researchers in various projects.

3. To create and implement training programs that enhance research skills and methodologies.

4. To establish collaborative spaces and networking opportunities for interdisciplinary research and innovation.

5. To Integrate cutting-edge technologies and resources to support diverse research needs.

### 4. Statement of Problem

Modern libraries face many problems, coupled with declining physical attendance, struggling with online information resources, and changing user expectations in the digital age. This challenge results from a gap between traditional library services and the dynamic needs of modern users. Many libraries struggle to adapt to innovative learning hubs, because they still apply traditional library services. There is a dire need to discover and understand the creative strategies employed by successful libraries to close this gap and transform them into active centers of engagement. Therefore, this study examines how libraries worldwide interpret and implement innovative practices in service delivery to inspire and empower users to make informed choices, presenting findings and recommendations on how libraries in the 21st century can be relevant to their core business of effectively supporting learning and community development.

## 5. Methodology

This qualitative research attempts to explore the initiatives undertaken by libraries across the globe to position themselves as learning centers for the empowerment of users through their myriad innovative services. This study relies on a critical review of a vast range of existing datasets available online or in any other form. These were mainly public documents constituting library websites, online reports, program descriptions, and promotional materials describing current innovative practices and strategies that libraries have adopted for user engagement. Such practices are contextualized through a critical literature review of broader trends and theoretical frameworks in library science. Thematic analysis allowed for the identification and interpretation of key themes and patterns across the data collected and from the literature. This approach allows one to examine how to apply various innovative strategies in implementing practical libraries, and thus, the impact on changing user empowerment. It also recognizes these possible limitations, which, by relying on online data and subjective qualitative analysis, are minimized through a systematic review of often cross-referenced sources to further increase the reliability and depth of the findings obtained.

## 6. Empowering Through Digital Resources and Tools

The introduction of online learning platforms, such as SWAYAM or Coursera, by Indian libraries can be used to further engage users. This embedding of space for online learning is carried out with the help of technical support, complementary workshops, and resource guides to establish and publicize these resources. Libraries disseminate platforms by scheduling numerous sessions, developing resource guides, and promoting recognition programs for completed courses. Access to a wide range of courses through the online library user community helps to create a community of lifelong learners. Encourage peer learning through study groups and forums. Students can also use and share resources or tools between themselves. This kind of integration of a wider array of library educational services situates the library permanently as an essential learning center for digital literacy and other skills within the community. Such initiatives are user-empowering, as they gain access to quality educational resources while underpinning their learning journey, thereby modernizing the role of the library in this contemporary educational landscape. (Gulati et al., 2021)

Libraries can do much to improve user engagement by developing incubation centers that stimulate innovation and collaboration. These centers turn the library into a lively, creative hub with collaborative workspaces, access to advanced tools, such as 3D printers, and resources for project development. Workshops, collaborations with other local organizations, and mentorship programs allow libraries to offer users poor learning experiences and professional connections. It works out that the organization of innovation challenges and promoting user-led initiatives work to empower patrons to become active problem-solvers and participants in community projects. Successful examples, such as the DeLaMare Science and Engineering Library and Arizona State University Libraries, have shown how incubation centers can support entrepreneurship and creativity. It not only does not engage users more with library resources, but also brings to the fore its powerful role in catalyzing innovation and community development through libraries redefined for the digital age. (Li et al., 2009)

Libraries can engage users in a highly enriching manner through diverse programming, catering to different interests and educational needs. With dynamic workshops in research skills, digital literacy, and emerging technologies, libraries can equip users with relevant competencies for the modern information landscape. Cultural events and book clubs help foster communities and spur intellectual discourse, while reading challenges motivate the continuous learning and exploration of library resources. Such programming, in collaboration with faculty to incorporate library resources into academic coursework, will not only enrich the educational experience but also reposition the library within the fabric of the learning process.

Programming at this level transforms libraries into places where every user is not a passive consumer of information but an engaged actor in a lively knowledge ecosystem. Through such activities, libraries increase usage but empower users to grow as individuals, develop skills, and connect with their communities, thereby entrenching the role of the library as an effective center for lifelong learning and innovation. (*User Engagement | University Libraries*, 2024)

The segmentation of users will become a key strategy toward increasing engagement in libraries, as this will help them personalize services and resources in accordance with the great and different needs of the users. This will happen by grouping users in terms of demographics, interests, and behaviors so that a library can come up with relevant offerings that would resonate with a particular group of people. This allows for the curation of collections, services, and programming that are cantankerous of the preferences of particular segments in the user population, such as digital resources for tech-savvy users or printed traditional materials for other lots. In addition, segmentation enables focused marketing strategies, in which the most interested audience receives relevant information about the activity. Such a personalized approach to library service provision not only makes services more relevant but also creates closer contact between users and the library. By identifying and serving distinctive needs in each group of users, libraries can take rates of engagement and participation in programs to a remarkable degree, thereby diffusing empowerment among users through the provision of resources and services that closely match their interests and requirements. (Zou et al., 2015) Recently, libraries have increasingly evolved into dynamic spaces that go beyond traditional roles in embracing digital resources and tools, diverse programs and events, interactive spaces and services, and community-led activities.

Libraries that assume the role of centers of research and innovation are core facilities for creating user engagement and empowering communities. This study considers how libraries capitalize on these key areas to create more inclusive, interactive, and innovative settings that meet the shifting needs of users. Such strategies that this study would adopt offer an indication of the future direction of library services and their roles in community development. (Zulkifli & Wahid, 2024)

**6.1 Digital resources and tools:** E-books, audio books, and access to online learning platforms such as NPTEL and Coursera are being developed as digital resources. Libraries, in this way, integrate those resources and provide the best access to knowledge and learning. This approach enriches user engagement because patrons can access a wide range of subjects and skills at their own pace.

**6.2 Programs and Events:** Participative programs in libraries bring in and retain users. Workshops, seminars, cultural programs, and guest speaker panels are interactive experiences that can serve a wide range of interests to help build community involvement. Such events add educational value while affording opportunities for social interaction and intellectual exchange.

**6.3 Interactive Spaces and Services:** Library design is rapid and takes the shape of creativity and collaboration. Examples of how libraries support experimentation, creation, and collaboration by their users include makerspaces, game areas, and flexible workspaces. These rewards actively participate in hands-on learning by using physical tools.

**6.4 Research and Innovation Centers:** In addition to realizing the role they play in supporting research at both academic and professional levels, libraries are also establishing specific centers to introduce advanced resources, personal support, and an atmosphere well organized for the purpose of research. Second, innovation and outstanding achievements in the research community are fostered in the library through the application of a reward system for local innovation.

**6.5 Community-led Activities:** Libraries increasingly engage in community activities through local talent shows, workshops on sharing skills, and sustainability initiatives. Such programmes bind communities and cater to local needs by fostering shared learning and mutual support.

**6.6 Philosophy and Framework**

With the philosophy of *"User First, Library Second, and Self Last"* at the heart of modern library services, the best practice for enhancing user engagement is delivered. It considers institutional operations and personal preferences only after the needs and experiences of a library's users. This is in line with S.R. Ranganathan's five laws of library science within this user-oriented framework (S R Ranganathan, 1931)

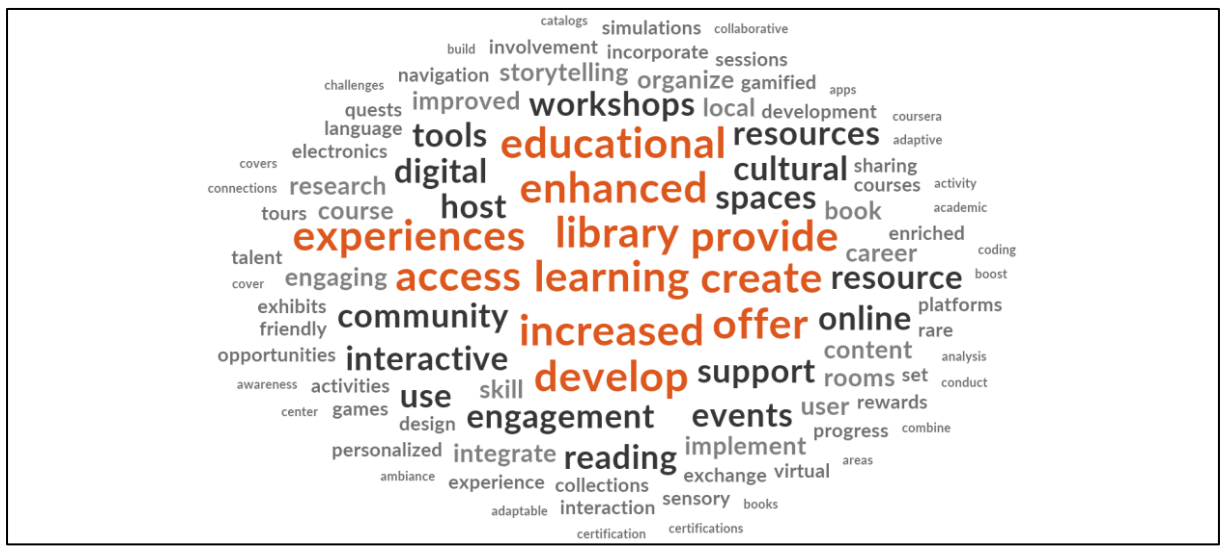

**Figure 1: Word Cloud of Library Engagement Topics**

This figure illustrates the main themes and focus areas of library user engagement. This word cloud visualizes the key terms related to library engagement strategies. Prominent words include "engagement," "libraries," "users," "social media," "tweets," and "community."

## 7. Data Analysis and Interpretation

**Information is primarily gathered using two methods.**

**7.1 Comprehensive Literature Review:** The Academic journals and conference proceedings on library science and innovation present theoretical and contextual background information.

**7.2 Online Database Search:** This involves an academic database and search engine search to collect studies, reports, articles, and other materials that help enrich the research. This provides further insight into how innovative strategies are implemented in libraries to empower users by identifying trends across data and the collected literature.

## 7.3 Software Tool Used to Analyse Data

NVivo software was used to organize and analyze the qualitative data, systematize coding, and develop themes.

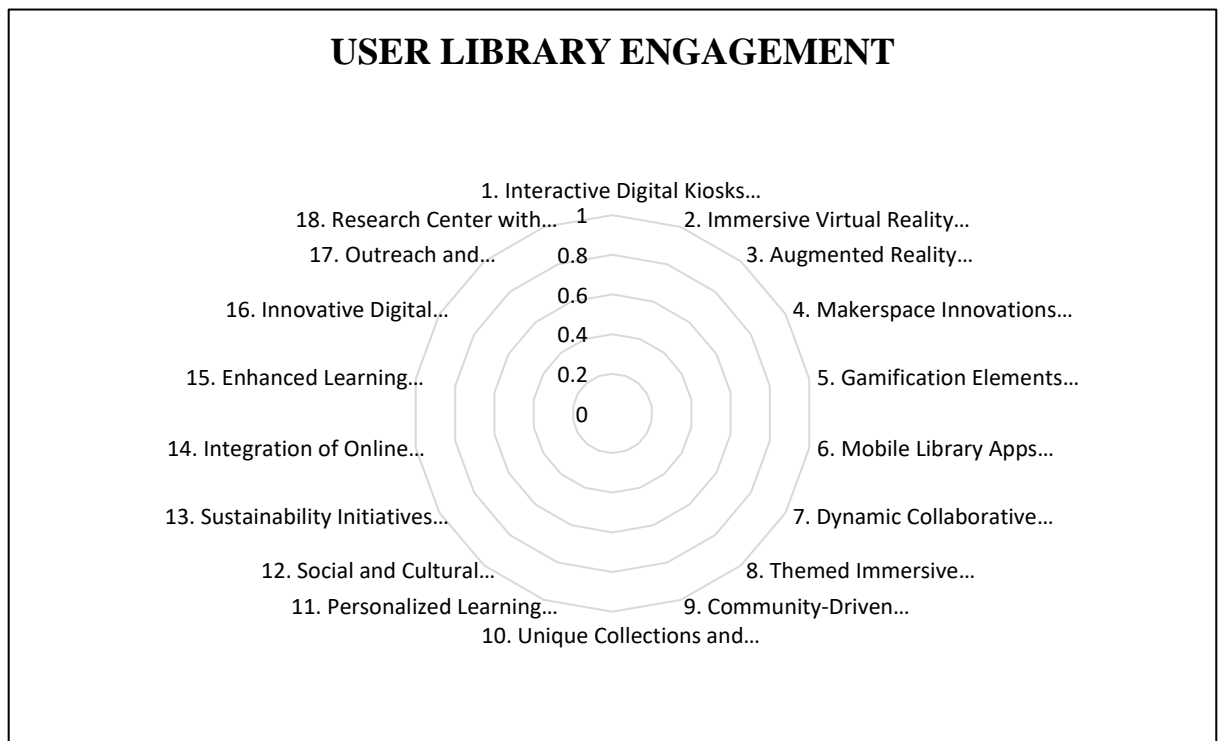

**Figure 3: Radar Chart of Modern Library Innovation Features**

This radar chart represents 18 of the most important features or aspects of modern library innovations. Different spokes describe another factor that varies from technological advancement to community engagement activities.

## 8. Key Findings on Effective Library Engagement Strategies

Below are the corresponding findings of the study on the effectiveness of different strategies for making library services better and more engaging with library users.

**8.1 Digital Resources:** In some cases, the feeling of a positive impact on the integration of digital resources concerning time and technology must change the libraries themselves. Providing a variety of digitally accessible tools through libraries will help to meet the changing needs of users and position them as modern institutions. (Mittal, 2017)

**8.2 Programming and Events:** The success of engaging in programs and events reveals that interactive and educational experiences are substantial in attracting and retaining users. Through the diversification of event offerings and community interactions, clients remain loyal to the institution, a component of lifelong learning.(Lubanga & Mumba, 2021)

**8.3 Interactive Spaces:** Building interactive spaces trends together with experiential learning and trends toward collaborative workspaces. For the library, designing interactive spaces that support creativity and teamwork creates a context for thoughts that might be outside the normal expectations. The bar is raised for libraries as vibrant and engaging places. (Lubanga & Mumba, 2021)

**8.4 Research and Innovation Centers:** Building research centers with systems of rewards demonstrates a strategic direction that focuses on the benefits accruing to academic growth. The presence of rewards for innovations and specialized resources in libraries helps to improve

the quality of scholarly work and, thus, offers support to researchers. (Lubanga & Mumba, 2021)

**8.5 Incubation Centers:** These runaway successes highlight the necessity to focus support specifically on start-ups and entrepreneurs. Libraries foster innovation and business through co-working spaces, mentorship, and opportunities to network with their staff, with the resultant effects reaching the local economy and community. (Ademilua, 2024). 276)

**8.6 Community Initiatives:** The effectiveness of community initiatives also shows that the importance of library community engagement goes beyond simply addressing the immediate needs of individuals or specific groups. Building the capacity to address community needs and enlightening common learning builds a better relationship with users and a comparison of diverse interests, thus contributing to local development. Libraries can effectively enhance this through multifaceted user engagement and foster innovation by diversifying the services offered. From digital resources and programs to interactive spaces, research, and incubation centers to community-driven initiatives, a mix of elements is combined in the library to realize a dynamic impact to match users' varied requirements and offer continuous support for creativity and academic excellence.(Ademilua, 2024). 276)

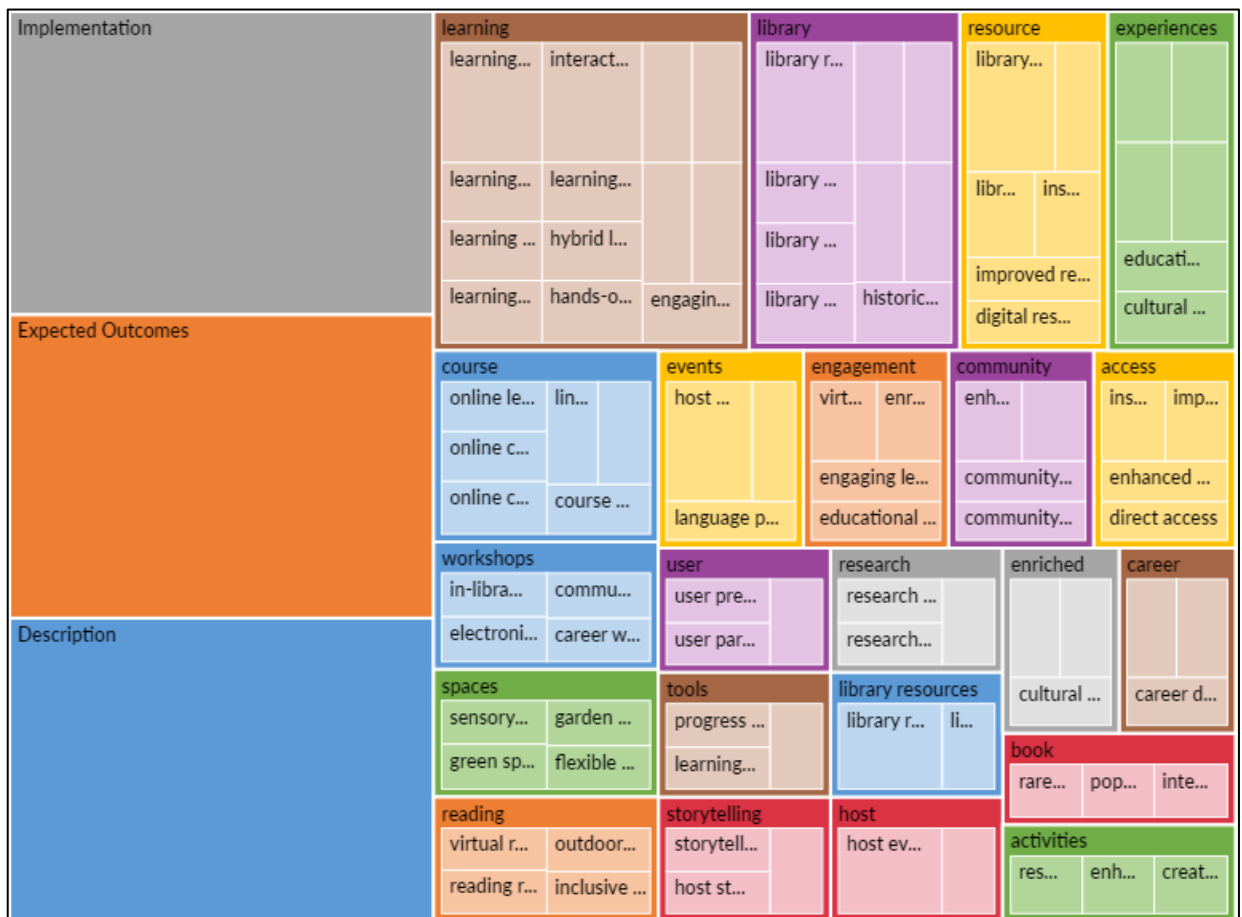

**Figure 4: Treemap of Library Information Systems and User Engagement Concepts**

This treemap indicates a hierarchical view of library information systems, user engagement, and related services. It shows the comparative importance and relationship of various aspects such as recommender systems, user profiling, and library services.

## 9. Limitations

The limitations of this study are as follows. This study provides important insights into the formulation of creative engagement strategies for the empowerment of library users. No empirical data from diverse library settings exist within it, though; that is because of the scope of literature reviewed and suspected biases in selected studies. Additional studies are required to replicate these findings in a broader context, as well as user demographics.

**9.1 Future Research Directions**: This approach expands the potential of future research on how the long-range effects of innovative library strategies impact user engagement and community building. This indicates the potential of a library to be more connected to its users through personalized services and continuous resourcing. New technologies such as AI and VR should be tried in the library to make offerings more personalized and interactive. Simultaneously, research design approaches that focus on user needs will provide insights into tailoring library services to respond to specific needs within distinct groups of users, thereby enhancing engagement. Research on community partnerships should be conducted to create an energetic and inclusive atmosphere within libraries. Finally, it evaluated the effectiveness of these creative strategies in terms of user satisfaction, learning outcomes, and the general use of libraries. (Safikhani et al., 2024)

## 10. Conclusion

This research registers a paradigm shift in library services toward an increasing emphasis on user-centered approaches and technological integration in modern libraries. Our findings show that successful libraries are constantly changing to dynamic, multifaceted learning hubs with a focus on user empowerment and engagement through new innovative strategies. Integrated into libraries are state-of-the-art technologies, such as Artificial Intelligence and personalization systems, which make user service much easier and more convenient. Simultaneously, with the introduction of interactive spaces, including makerspaces and incubation centers, it began gathering momentum toward becoming a mature community of creativity, collaboration, and innovation. Effective segmentation mechanisms allow it to create more target-oriented services for the manifold needs of its diverse patronage and generate much deeper engagement. These strategies hold enormous potential to address challenges specific to rural and underserved communities, including the role libraries may play in furthering digital divides and supporting sustainable development. Second, after years of varied programming, online learning platforms have become the key components of lifelong learning and community engagement. However, challenges related to resource allocation, staff training, and balancing innovative digital-based initiatives with more traditional values within the library remain. Thus, in a nutshell, the ability to adapt constantly and innovate further will be the key to the future success of libraries while retaining their core mission of democratizing access to knowledge. The empowerment of users through rich experiences facilitated by creative engagement strategies is a transformative role of libraries. Libraries thus reinforce their significance and scale up their influence as institutions for learning, creativity, and community building. These findings will be useful in informing the practice of library professionals, policymakers, and community leaders in guiding the transformation of libraries into dynamic agencies equipped with the skills necessary for responding to emerging societal challenges. (Proffitt et al., 2015)

## 11. Suggestions & Recommendations to Practice in Indian Libraries

### 11.1 Leverage Data Analytics

Collected data on circulation, attendance at programs and events, and user interactions to gauge users' preferences and behavior. These are then used to design library programs and allocate resources that support user interests. (Tešendić & Boberić Krstićev, 2019)

### 11.2 Implementing a Rewards System

This can also provide an additional layer to track usage to find the most active library users who may receive special rewards in the form of gift cards, special access, discounts on value-added services, etc. Gamification can easily be introduced through monthly challenges that give special rewards to users who visit more frequently. (Giannaros et al., 2020)

**11.2.1 Promote and Refine:**

Publicize the reward program through social media, newsletters, and posters in libraries. Gather user feedback to make adjustments in rewards and add enrichment to the program based on the participants' needs. (Giannaros et al., 2020)

**11.2.2 Showcase Impact:**

This will be useful in demonstrating how reward programs change with the increased use of the library, as well as community involvement. The success of the program will be highlighted in order to provide grounding for any future funding and continued development. This is one strategy that will help win library patrons for a lively library community and retain them. (Giannaros et al., 2020)

# References

Ademilua, S. (2024). *Promoting Competitive Intelligence Strategies to Advance Library… | 276. 4 (2)*, 275–296. https://doi.org/10.17509/ijomr.v4i2.75300

Giannaros, A., Kotsopoulos, K., Tsolis, D., & Pavlidis, G. (2020). *Creating a Personalised Experience for Libraries' Visitors* (pp. 491–498). https://doi.org/10.1007/978-3-030-36126-6_55

Gulati, S., Sharma, R., Scholar, A. K. R., & Chakravarty, R. (2021). *Understanding User Perceptive and Satisfaction Level towards MOOCs: A Comparative analysis of SWAYAM and Coursera*.

Henkel, M., Ilhan, A., Mainka, A., & Stock, W. (2018). *Open Innovation in Libraries*. Hawaii International Conference on System Sciences. https://doi.org/10.24251/HICSS.2018.522

Huang, H., Chu, S., Liu, L., & Zheng, P. (2017). Understanding User-Librarian Interaction Types in Academic Library Microblogging: A Comparison Study in Twitter and Weibo. *The Journal of Academic Librarianship*, *43*. https://doi.org/10.1016/j.acalib.2017.06.002

Li, X., Wang, Y., Fu, L., & Xu, M. (2009). The university library: Incubation center of research innovation literacy. *The Electronic Library*, *27*, 588–600. https://doi.org/10.1108/02640470910979552

Lubanga, S., & Mumba, J. (2021). Research and Development (R&D), Creativity and Innovation in Academic Libraries in Malawi: A Way To Rethink Library Development in the 21st Century. *SSRN Electronic Journal*. https://doi.org/10.2139/ssrn.3867430

Mittal, A. (2017). *User Behavior towards Digital Resources in Mohinder Singh Randhawa Library: A Study*. *4*(2).

Otike, F., Barát, Á. H., & Kiszl, P. (2022). Innovation strategies in academic libraries using business entrepreneurial theories: Analysis of competing values framework and disruptive innovation theory. *The Journal of Academic Librarianship*, *48*(4), 102537. https://doi.org/10.1016/j.acalib.2022.102537

Proffitt, M., Michalko, J., & Renspie, M. (2015). *Shaping the Library to the Life of the User: Adapting, Empowering, Partnering, Engaging*.

S R Ranganathan. (1931). *The Five Laws Of Library Science*. http://archive.org/details/in.ernet.dli.2015.283188

Safikhani, S., Gross, B., & Pirker, J. (2024). *The Application of Procedurally Generated Libraries in Immersive Virtual Reality* (arXiv:2406.01128). arXiv. http://arxiv.org/abs/2406.01128

Tešendić, D., & Boberić Krstićev, D. (2019). Business Intelligence in the Service of Libraries. *Information Technology and Libraries*, *38*(4), 98–113. https://doi.org/10.6017/ital.v38i4.10599

*User Engagement | University Libraries*. (2024). https://library.louisiana.edu/services/user-engagement

Zou, H., Chen, H., & dey, S. (2015). Exploring User Engagement Strategies and Their Impacts with Social Media Mining: The Case of Public Libraries. *Journal of Management Analytics*, *2*, 1–19. https://doi.org/10.1080/23270012.2015.1100969

Zou, H., Chen, H. M., & Dey, S. (2020). *Engaging Users through Social Media in Public Libraries*.

Zulkifli, A. F., & Wahid, E. E. A. (2024). *The Impact of Academic Digital Library Engagement Among Postgraduate Scholars*. *14*(1).